\documentclass[showpacs,preprintnumbers,amsmath,amssymb,12pt,floatfix,epsfig]{revtex4}
\usepackage{graphicx}

\begin{document}
\draft
\title{Signals for strange quark contributions to the neutrino (antineutrino)
scattering in quasi-elastic region }
\author{ Myung-Ki Cheoun$^{1)}$, K. S. Kim$^{2)}$
\footnote{Corresponding author : kyungsik@hau.ac.kr}}
\address{1)Department of Physics,
Soongsil University, Seoul, 156-743, Korea \\ 2)School of Liberal
Arts and Science, Korea Aerospace University, Koyang 412-791,
Korea }

\begin{abstract}
Strange quark contributions to the neutrino (antineutrino)
scattering are investigated on the elastic neutrino-nucleon
scattering and the neutrino-nucleus scattering for $^{12}$C target
in the quasi-elastic region on the incident energy of 500 MeV,
within the framework of a relativistic single particle model. For
the neutrino-nucleus scattering, the effects of final state
interaction for the knocked-out nucleon are included by a
relativistic optical potential. In the cross sections we found
some cancellations of the strange quark contributions between the
knocked-out protons and neutrons. Consequently, the asymmetries
between the incident neutrino and antineutrino which is the ratio
of neutral current to charged current, and the difference between
the asymmetries are shown to be able to yield more feasible
quantities for the strangeness effects. In order to explicitly
display importance of the cancellations, results of the exclusive
reaction $^{16} O(\nu , \nu^{'} p)$ are additionally presented for
detecting the strangeness effects.
\end{abstract}
\narrowtext
\maketitle


Since the exploration of a spin structure of proton at the EMC
measurement \cite{emc89}, the deep-inelastic scattering of leptons
from nucleons has played important roles for studying the
distribution of quarks and gluons. In the quark parton model for
the nucleon, the cross section of the scattering with polarized
leptons on polarized nucleons is usually given by four structure
functions, $F_1 (x), F_2 (x), g_1 (x)$, and $ g_2 (x)$
\begin{equation}
F_2 (x) = x {\mathop{\Sigma}_{q}} e_q^2  q (x) = 2 x F_1 (x)~,~
g_1 (x) = { 1 \over 2} {\mathop{\Sigma}_{q}} e_q^2  \Delta q (x)
\end{equation}
with $x = Q^2 / 2 p \cdot q$ and $Q^2 = - {( p^{'} - p )}^2$ for
the momentum $p$ of the initial nucleon and the momentum transfer
$q$ from leptons. $e_q , q (x)$, and $ \Delta q(x)$ denote a
charge, a parton distribution and a polarized parton distribution
function for quarks flavor $q$, respectively \cite{Albe02,garvey}.
In experimental aspect, the structure functions are usually
measured as their integral forms because of their $Q^2$ dependence
\begin{equation}
\Gamma_1 (Q^2) = \int_0^1 g_1 ( x, Q^2 ) dx~.
\end{equation}
Available experimental data for the $\Gamma_1 (Q^2 )$ are located
in $0.115 \sim 0.126$ depending on the region, $ 3 \le Q^2 \le
10.7 $ \cite{Albe02}. If $\Delta q$ is defined as a difference of
the total numbers of quarks and antiquarks in the nucleon with a
helicity equal and opposite to the spin of the nucleon
\begin{equation}
\Delta q  = \int_0^1 {\mathop{\Sigma}_{r = \pm 1}} r [ q^{(r)} (x)
+ {\bar q}^{(r)} (x) ] dx  ~,
\end{equation}
$\Delta q$ stands for the contribution of $q$-quarks and ${\bar
q}$-antiquarks to the spin of the nucleon. For example, in the
infinite momentum frame (quark-parton model), $\Gamma_1 ( Q^2)$ is
evaluated in terms of the $\Delta q$ with some constraints among
them
\begin{equation}
\Gamma_1^p = { 1 \over 2}  ( { 4 \over 9} \Delta u + { 1 \over 9}
\Delta d + {  1 \over 9} \Delta s)~,~ \Delta u + \Delta d - 2
\Delta s = 3 F - D~,~ \Delta u - \Delta d = g_A
\end{equation}
with $D$ and $F$ constants in SU(3) symmetry.

On the other hand, the axial form factors $G_A^q ( Q^2)$ and the
axial charges $g_A^q$ are given by the matrix elements and the
diagonal matrix elements of the axial current
\begin{equation}
< p^{'} | {\bar q} \gamma_{\mu} \gamma_5 q | p > = {\bar u} (
p^{'}) \gamma_{\mu} \gamma_5 G_A^q ( Q^2 ) u (p)~,~< p | {\bar q}
\gamma_{\mu} \gamma_5 q | p > = 2 M s_{\mu} g_A^q~,
\end{equation}
where $M$ and $s_{\mu}$ are respectively the mass and spin vector
of the nucleon. In particular, for our primary interest, the axial
coupling constant $g_A^s$ (or axial charge) induced by strangeness
quark turned out to be equal to the $\Delta s$, which is related
to the nucleon spin by the $s$ quark contribution. Though its
value extracted from both experimental and/or theoretical results
has not been fixed yet, several groups reported the value as $0.0
$ \cite{Young06}, $-0.08 \pm 0.06, -0.12 \pm 0.06 $ \cite{Pate05}
and $-0.18 \pm 0.05 $ \cite{emc89}. Therefore, its value is
allowed to vary in $0 < g_A^s < - 0.19$ for further discussions of
its effects in the elastic neutrino (antineutrino) $(\nu {(\bar
\nu}))$ scattering.

In this paper, although the strangeness effects in vectorial parts
can be studied by parity not only violating (or polarized electron
scattering) but $\nu$ scattering, we focus on the strangeness
$g_A^s$ effect in the axial part by studying the $\nu {(\bar
\nu})$ scattering on the quasi-elastic (QE) region, where
inelastic processes like pion production and delta resonance are
excluded.

Before going further, we briefly summarize the present status for
signals of the strangeness effects in the $\nu {(\bar \nu})$
scattering. Since the vectorial parts are rarely contributed due
to the given Weinberg angle in the reaction, the elastic cross
section on the proton $\sigma ( \nu p \rightarrow \nu p)$ mediated
by a neutral current (NC) reaction is sensitive mainly on the
$g_A^s$ value. But the measurement of the cross section is not
experimentally easy, so that one usually resorts to the ratio of
proton to neutron cross sections $R_{p/n} = { {\sigma ( \nu p
\rightarrow \nu p ) } / { \sigma (\nu n \rightarrow \nu n )}}$
\cite{garvey}. The measurement of this ratio has also some
difficulties in the neutron detection. Since the charged current
(CC) cross section is relatively insensitive to the strangeness,
the ratio $R_{NC/CC} = { {\sigma ( \nu p \rightarrow \nu p ) } / {
\sigma (\nu n \rightarrow \mu^- p )}}$ is suggested as a plausible
signal for the nucleon strangeness although the signal goes down
by a factor 2 \cite{Vent06}. In this paper, more precise analysis
for those quantities are carried out in searching other feasible
quantities for the strangeness effects in the $\nu {(\bar
\nu})$-scattering.

The elastic neutrino-nucleon $(\nu {(\bar \nu})- N)$ cross section
by the NC reaction is expressed in terms of the Sachs (vector and
axial) form factors \cite{Albe02,cheoun08}
\begin{eqnarray}
{( {  {d \sigma  } \over {d Q^2  }} )}_{\nu ({\bar \nu}   )
}^{NC}&=& { {G_F^2 } \over {2 \pi }} [ { 1 \over 2} y^2 { (G_M^V
)}^2 + ( 1 - y - { { M }\over {2 E_{\nu} }} y ) { {{ (G_E^V )}^2 +
{ E_{\nu} \over 2 M} y { (G_M^V )}^2 } \over {1 + {  E_{\nu} \over
2 M} y }} \nonumber \\ && + ( { 1 \over 2} y^2 + 1 - y + { { M }
\over {2 E_{\nu} }} y ){ (G_A )}^2 + h 2 y ( 1 - { 1 \over 2 } y )
G_M^V G_A ]~.
\end{eqnarray}
Here, $E_{\nu}$ is the incident $\nu ({\bar \nu})$ energy in the
laboratory frame, and $y = { { p \cdot q } / { p \cdot k }} = { {
Q^2 } / {2 p \cdot k }}$ with initial four momenta $k$ of $\nu
({\bar \nu})$, target nucleon $p$, and four momentum transfer $q$
to the nucleon, respectively. $G_F \simeq 1.16639 \times 10^{-11}$
MeV$^{-2}$ is the Fermi constant. Vectorial Sachs form factors are
written as electric and magnetic form factors of the nucleon
\begin{eqnarray}
G_{M,E}^{V,~ p(n)} ( Q^2) &=&({\frac 1 2} - 2 \sin^2 \theta_W )
G_{M,E}^{p(n)} ( Q^2) - {\frac 1 2} G_{M,E}^{n(p)}( Q^2) -{\frac 1
2} G_{M,E}^s ( Q^2)~~ \mbox{for~ NC}~ \nonumber \\& =& (
G_{M,E}^{p} ( Q^2) -
 G_{M,E}^{n}( Q^2))~~~~~~~~~~~~~~~~~~~~~~~~~~~~~~~~~~~~~~~~\mbox{for~ CC}~,
\nonumber \\ G_M^s (Q^2) &=& {  {Q^2 F_1^s + \mu_s  } \over { ( 1
+ \tau) {( 1 + Q^2/M_V^2)}^2 }}, ~G_E^s (Q^2) = {  {Q^2 F_1^s -
\mu_s \tau  } \over { ( 1 + \tau) {( 1 + Q^2/M_V^2)}^2 }}~,
\end{eqnarray}
where $\mu_s =G_M^s (0) $ is a strange magnetic moment. The axial
form factor is usually presumed as a dipole form
\begin{eqnarray}
G^{NC}_A (Q^2) &=&{\frac 1 2} (\mp g_A + g_A^s)/(1+Q^2/M_A^2)^2~~
\mbox{for~ NC}~ \nonumber \\G_A^{CC} (Q^2) & =& - g_A / {( 1 + Q^2
/ M_A^2)}^2~~~~~~~~~~~~~ \mbox{for~ CC}~, \label{gs}
\end{eqnarray}
where $g_A=1.262$ and $M_A$ are the axial coupling constant and
the axial cut off mass, respectively. $-(+)$ coming from the
isospin dependence denotes the knocked-out proton (neutron),
respectively. The $g_A^s $ appeared explicitly in Eq.(8)
represents the strange quark contents in the nucleon. Of course,
the strangeness effects could contribute to other form factors,
but their effects turned out to contribute only a few \% to those
physical quantities \cite{cheoun08}.

In order to calculate the $\nu ( {\bar \nu})$-nucleus $(\nu  - A)$
scattering, we choose the nucleus fixed frame where the target
nucleus is seated at the origin of the coordinate system. The
four-momenta of the incident and outgoing neutrinos
(antineutrinos), the target nucleus, the residual nucleus, and the
knocked-out nucleon are labelled $p_i^{\mu}$ and $p_f^{\mu}$,
$p_A^{\mu}$, $p_{A-1}^{\mu}$, and $p^{\mu}$, respectively. In the
laboratory frame, the inclusive cross section, which does not
detect the outgoing $\nu$ (${\bar \nu}$), is given by the
contraction between lepton and hadron tensor
\cite{kimplb,umino,udias,giusti1}
\begin{eqnarray}
{\frac {d\sigma} {dT_p}} &=& 4\pi^2{\frac {M_N M_{A-1}} {(2\pi)^3
M_A}} \int \sin \theta_l d\theta_l \int \sin \theta_p d\theta_p p
f^{-1}_{rec} \sigma^{Z,W^\pm}_M [ v_L R_L  + v_T R_T + h v'_T R'_T
]~, \label{cs}
\end{eqnarray}
where $\theta_l$ denotes the scattering angle of the lepton.
$\sigma^{Z,W^{\pm}}_M$ for NC and CC is defined by
\begin{equation}
\sigma^Z_M  =  {\left ( {\frac {G_F \cos (\theta_l/2) E_f M_Z^2}
{{\sqrt 2} \pi (Q^2 + M^2_Z)}} \right )}^2~~,~ \sigma^{W^\pm}_M  =
\sqrt{1 - {\frac {M^2_l} {E_f}}} \left ( {\frac {G_F \cos
(\theta_C) E_f M_W^2} {2\pi (Q^2 + M^2_W)}} \right )^2~~,
\end{equation}
where $M_Z$ and $M_W$ are the rest mass of $Z$-boson and
$W$-boson, respectively. $\theta_C$ denotes the Cabibbo angle
given by $\cos^2 {\theta_C} \simeq 0.9749$. The recoil factor
$f_{rec}$ is given as
\begin{equation}
f_{rec} = {\frac {E_{A-1}} {M_A}} \left | 1 + {\frac {E_p}
{E_{A-1}}} \left [ 1 - {\frac {{\bf q} \cdot {\bf p}} {p^2}}
\right ] \right |. \end{equation}
Nuclear response functions to the weak current are denoted as
$R_L, R_T$ and $R_T^{'}$, whose detailed forms are referred from
Ref. \cite{kimplb}.

\begin{figure}
\vskip-3.5cm
\includegraphics[width=0.5\linewidth]{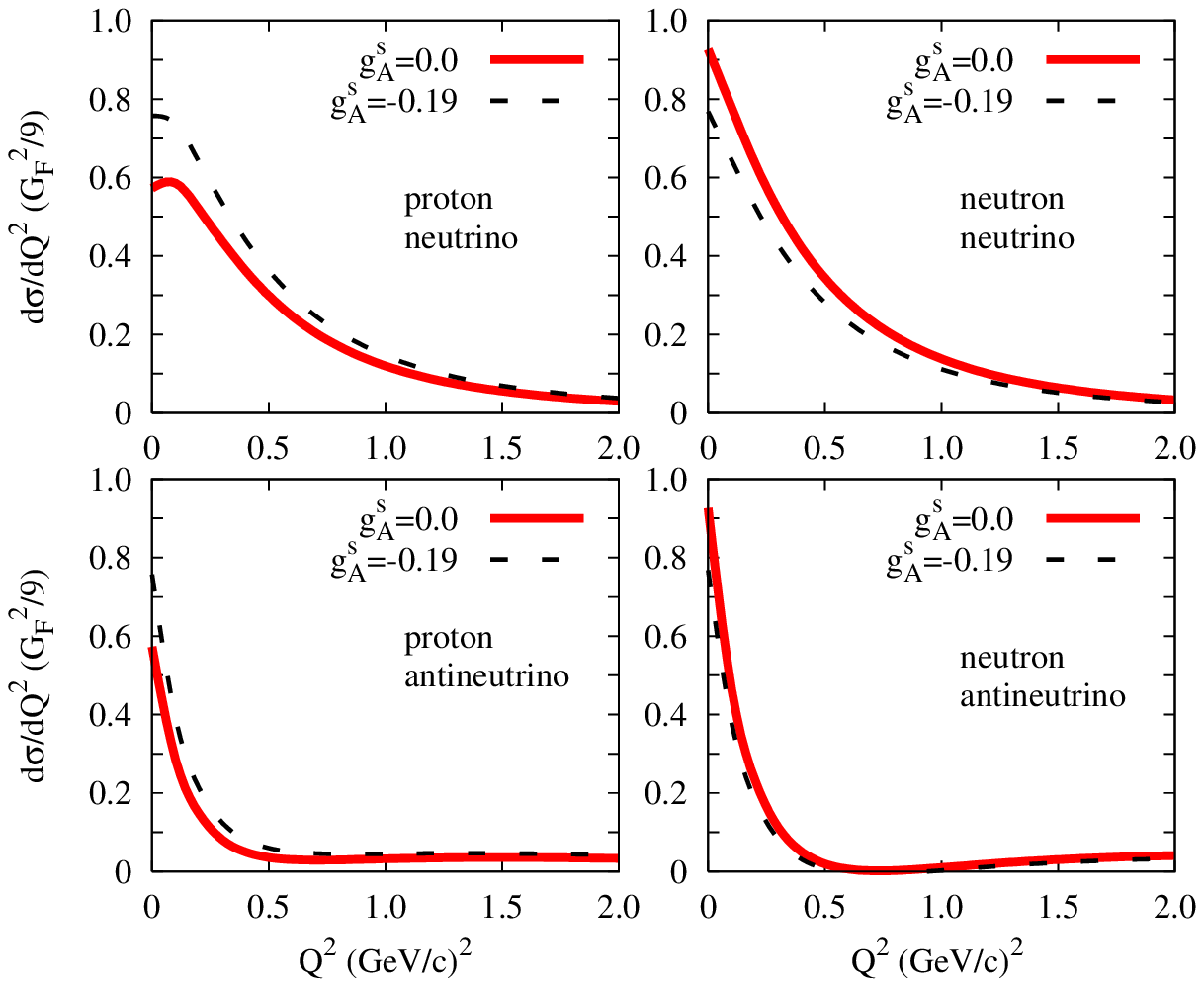}
\includegraphics[width=0.5\linewidth]{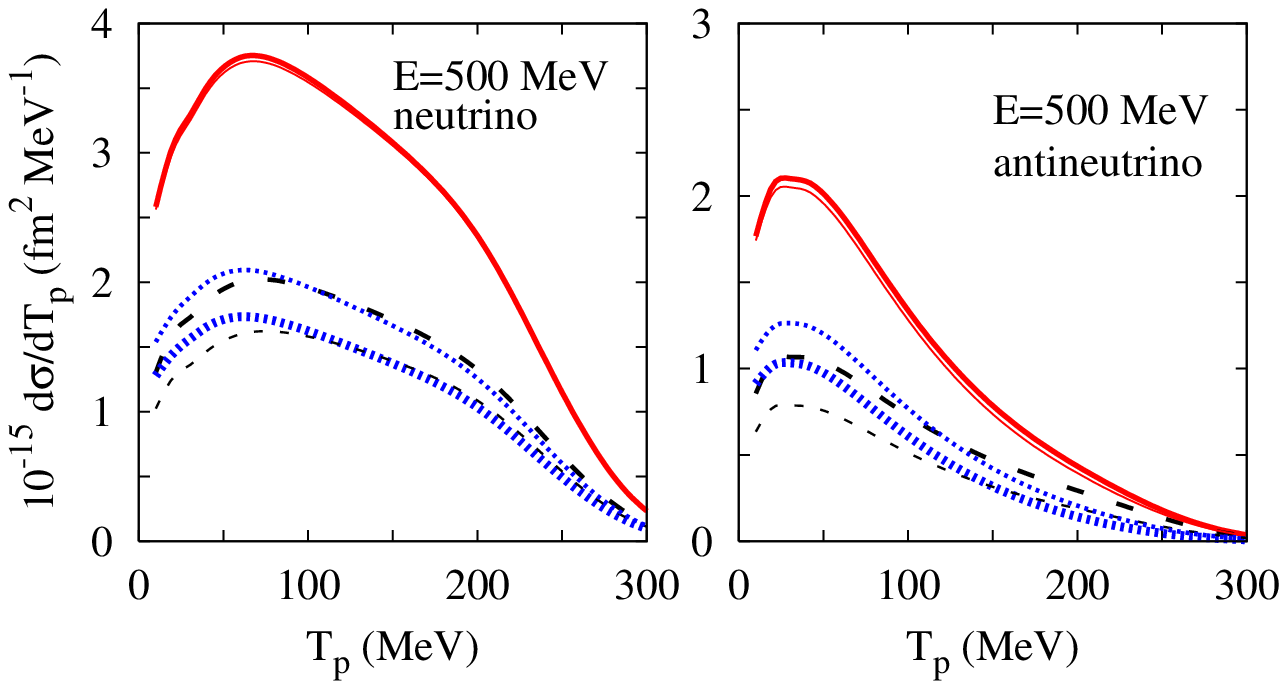}
\caption{(Color online) Upper 4 figures are differential cross
sections by the NC scattering on proton (left) and neutron
(right), Eq. (6), in a unit ($G_F^2/9 \simeq 6.05 \times 10^{-39}~
[{ {cm }^2/{(GeV/c) }^2}]$), as a function of $Q^2$ for $E_{\nu
({\bar \nu})}=500$ MeV. They are calculated for $g_A^s = -0.19$
and 0.0 cases, respectively. The lowest panel is for $^{12}$C($\nu
(\bar \nu), \nu'(\bar \nu)$) cross section, Eq.(9), as a function
of the knocked out nucleon kinetic energy $T_p$ at $E_{\nu ({\bar
\nu})}=500$ MeV, where left (right) panel is for the $\nu$ ($\bar
\nu$) scattering, respectively. Solid curves are the results for
the cross sections, dashed and dotted lines are the contributions
of the proton and the neutron, respectively. Thick and thin lines
are calculations with $g^s_A=-0.19$ and $g^s_A=0.0$.}
\label{cross}
\end{figure}

Detailed results of the cross sections on the proton and the
neutron target for $g_A^s = -0.19$ and $0.0$ are shown in the
upper parts of Fig. \ref{cross}. Cross sections by the incident
$\nu ({\bar \nu})$ on the proton are usually enhanced in the whole
$Q^2$ region by the $g_A^s$, 15\% maximally, while they are
reduced on the neutron. Other quantities related to these cross
sections, such as asymmetries and ratios between $\nu$ and ${\bar
\nu}$ in the NC reaction, also show similar effects
\cite{cheoun08}. Therefore the strangeness effect is not large
enough to be discernible in the cross sections, asymmetries, and
ratios if we recollect the persisting experimental error bars in
the BNL data \cite{garvey}.

The lowest part of Fig. \ref{cross} exhibits the strange quark
contributions to the cross section on $^{12}$C$( \nu ( {\bar \nu}
), \nu^{'} ( {\bar \nu^{'}} ))$ reaction. For directly comparing
with the nucleon case, we present them in terms of outgoing
nucleon kinetic energy $T_N$ because $Q^2 = 2 M T_P$ would hold on
the free nucleon inside nuclei. Thick and thin curves are the
results for $g^s_A=-0.19 $ and $ 0.0 $, respectively. Note that a
relativistic optical potential for the final nucleon is taken into
account for the FSI \cite{clark}. Detailed discussions about the
FSI are done at Ref. \cite{kim07}. Peak positions for $^{12}$C
nuclei, which does not show up in the nucleon case, just come from
the binding energy of nucleons inside nuclei.

To analyze cross sections (solid curves) for the $\nu -^{12}$C
scattering, we present separately each contribution via final
neutrons (dotted curves) and final protons (dashed curves) in the
figure. The effect of $g^s_A$ for the proton is increased by about
30\%, but for the neutron it is decreased by 30\%, maximally,
independently of the incident energy. These individual $g_A^s$
effects on each nucleon resemble exactly those of each nucleon in
the upper parts of Fig. \ref{cross}, that is, the $g_A^s$ effect
enhances the cross section by protons and decreases that by
neutrons.

However, total net $g_A^s$ effects severely depend on the
competition between the protons and neutrons processes because of
the summation of all nucleons. In the case of $^{12}$C, the
enhancement by the proton is nearly compensated by the neutron
process, so that the net effect increases the cross section only
below 3\% for the given $E_{\nu}$.

For the ${\bar \nu}-^{12}$C process, the $g_A^s$ effect enhances
the cross section by about 20\% on the protons, but reduces them
by 20\% the neutrons. The reduction by neutrons is nearly balanced
by the enhancement due to protons. Consequently, the net effect of
the strange quark reduces also the cross section only below 2\%
for the given $E_{\bar \nu}$, similarly to the $\nu$ case.

From these results, in the $\nu - A$ cross sections, the $g^s_A$
effect contributes more positively to the proton while it does
negatively to the neutron, and is cancelled out finally. In the
case of $^{12}$C, the resultant effects are only within a few \%
by the cancellation between the final protons and neutrons.
Therefore, the $g_A^s$ effect in nuclei is too small to be deduced
from the A$( \nu, \nu^{'})$ cross section.


On the other hand, the cross section for the CC scattering is
given with the following replacement into the NC cross sections of
Eqs. (6) and (9)
\begin{equation}
{( {  {d \sigma  } \over {d Q^2  }} )}_{\nu ({\bar \nu}   )
}^{CC}= {( {  {d \sigma  } \over {d Q^2  }} )}_{\nu ({\bar \nu} )
}^{NC} ( G_E^V \rightarrow G_E^{CC}, G_M^V \rightarrow G_M^{CC},
G_A \rightarrow G_A^{CC})~,
\end{equation}
with $ G_{E}^{CC} = G_E^p (Q^2) - G_E^n (Q^2)~,~G_{M}^{CC} = G_M^p
(Q^2) - G_M^n (Q^2) $. Since the relevant form factors are
expressed only by the electro-magnetic form factors the CC
scattering is not nearly influenced by the strangeness in the form
factors. Therefore, the ratios of the NC and CC reactions given by
\begin{equation}
R_{NC/CC}= {\frac {\sigma (\nu, \nu' p)} {\sigma (\nu, \mu^- p)}}
= {\frac {\sigma_{NC}^{\nu p}} {\sigma_{CC}^{\nu }}}~, {\bar
R_{NC/CC}} = {\frac {\sigma ({\bar \nu}, {\bar \nu}' n)} {\sigma
({\bar \nu}, \mu^+ n)}} = {\frac {\sigma_{NC}^{{\bar \nu}n}}
{\sigma_{CC}^{{\bar \nu}}}}~, \label{ratio}
\end{equation}
have been suggested for probing the strangeness on the nucleon or
nuclei. Moreover any possible nuclear structure effects in these
reactions are expected to be cancelled out. Since these ratios are
focused on the knocked-out nucleon, we introduce another
definition of the ratios by focusing on the nucleon inside the
target nucleus as follows
\begin{equation}
R'_{NC/CC}= {\frac {\sigma (\nu, \nu' n)} {\sigma (\nu, \mu^- p)}}
= {\frac {\sigma_{NC}^{\nu n}} {\sigma_{CC}^{\nu }}}~,~ {\bar
R}'_{NC/CC} = {\frac {\sigma ({\bar \nu}, {\bar \nu}' p)} {\sigma
({\bar \nu}, \mu^+ n)}} = {\frac {\sigma_{NC}^{{\bar \nu}p}}
{\sigma_{CC}^{{\bar \nu}}}}. \label{ratio-ch}
\end{equation}
Since the charge exchange is not included, these $R^{'}$ and
${\bar R}^{'}$ quantities have the same meaning as $R_{NC/CC} =
{\sigma (\nu n \rightarrow \nu n)}/{\sigma (\nu n \rightarrow
\mu^- p)}, {\bar R}_{NC/CC} = {\sigma ({\bar \nu} p \rightarrow
{\bar \nu} p)}/{\sigma ({\bar \nu} p \rightarrow {\mu^+} n)}$ on
nucleon level, namely, $R^{'}$ and ${\bar R}^{'}$ mean the ratios
on the nucleon inside nuclei bombarded by incident $\nu ({\bar
\nu})$. Results and relevant discussions for the ratios on the
$^{12}$C target are given at Fig. \ref{ratio}. The strangeness
effects are not so large than those of the cross sections at Fig.
\ref{cross}. Divergence of high $T_P$ is associated with the
effect of outgoing lepton mass in the CC reaction \cite{Albe02}.


In order to find more feasible $g_A^s$ signals, the difference of
the cross sections between incident $\nu$ and ${\bar \nu}$ is
suggested as a candidate for the effects\cite{Albe02,kimplb}.
Since $h=-1$ $(h=+1)$ in Eqs.(6) and (9) corresponds to the
helicity of the incident $\nu$ (${\bar \nu}$), a difference of the
cross sections is summarized as a simple form. For instance, for
the $\nu - N$ scattering, the difference of the cross section is
given as
\begin{equation}
{( {  {d \sigma  } \over {d Q^2  }} )}_{\nu     }^{NC} - {( { {d
\sigma  } \over {d Q^2  }} )}_{ {\bar \nu}    }^{NC} = - { {G_F^2
} \over { 2 \pi }}~ 4 y ( 1 - { 1 \over 2} y ) G_M^V G_A~.
\end{equation}
Similar conjectures can be done for the $\nu - A$ scattering from
Eq. (9).

Since the difference of the cross sections between $\nu$ and
${\bar \nu}$ can be expressed only by a few form factors, more
efficient observable are possible if we consider asymmetries
between $\nu$ and ${\bar \nu}$ relative to the CC reactions.
Moreover it enables us to distinguish each strangeness
contribution in the two strangeness effects, {\it i.e.} vectorial
and axial strangeness parts at small $Q^2$ region
\begin{eqnarray}
A_{NC/CC}^p & = & { {\sigma_{NC}^{\nu p} - \sigma_{NC}^{{\bar \nu}
p} } \over {(\sigma_{CC}^{\nu n} - \sigma_{CC}^{{\bar \nu} p}  )}}
=  0.12 - 0.12 { {g_A^s } \over { g_A }}
- 0.13 { {G_M^s } \over {G_M^3  }}~,\\
\nonumber A_{NC/CC}^n & = & { {\sigma_{NC}^{\nu n} -
\sigma_{NC}^{{\bar \nu} n} } \over {(\sigma_{CC}^{\nu n} -
\sigma_{CC}^{{\bar \nu} p} )}}  = 0.16 + 0.16 { {g_A^s } \over {
g_A }} + 0.13 { {G_M^s } \over {G_M^3  }}~,
\end{eqnarray}
where the third terms with $G_M^s ( Q^2 = 0) = \mu_s$, $G_M^3 = {
(G_M^p - G_M^n}) / 2$, $\mu_p = 2.79$, and $\mu_n = -1.91$ come
from the vectorial part. Since $ \vert G_M^s/G_M^3 \vert $ and $
\vert g_A^s/g_A \vert$ are approximately 0.2, the strangeness
effects from the vector and axial parts are comparable, in
principle. It could be questioned if Eq.(16) still holds even in
the case of a nucleus. But, we assume that it holds because the
outgoing nucleon can be described as quasi-freely bombarded by the
incident $\nu ({\bar \nu})$.

\begin{figure}
\includegraphics[width=0.6\linewidth]{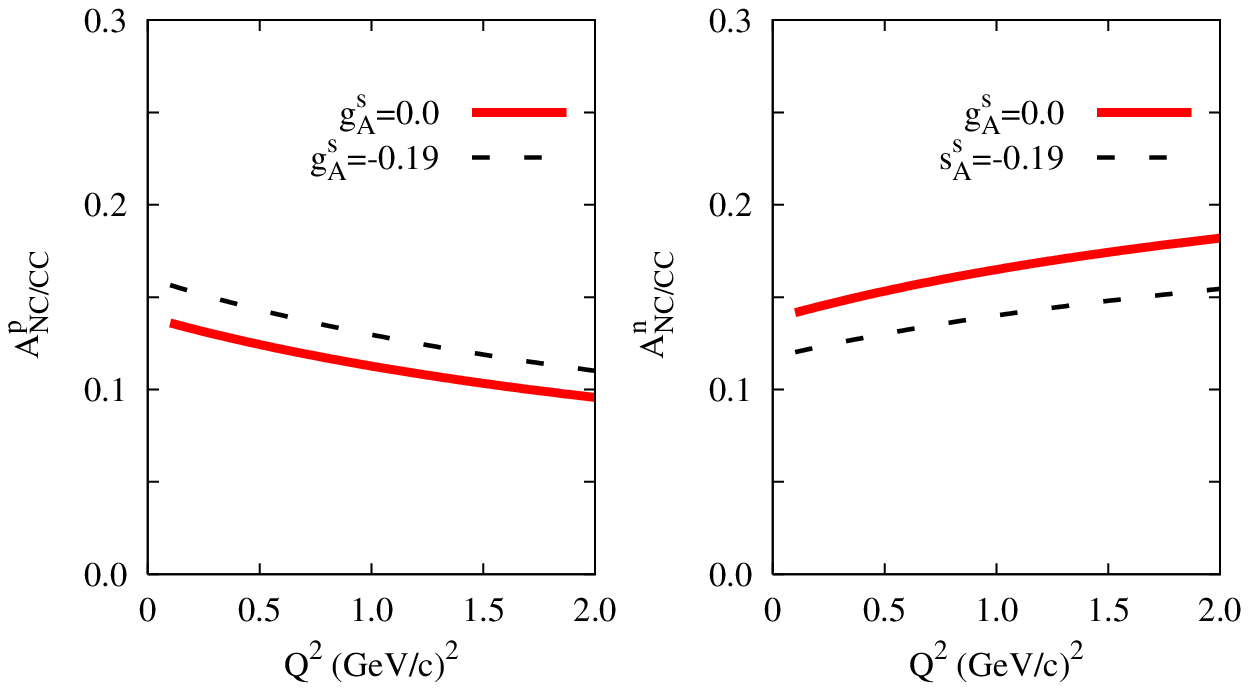}
\includegraphics[width=0.66\linewidth]{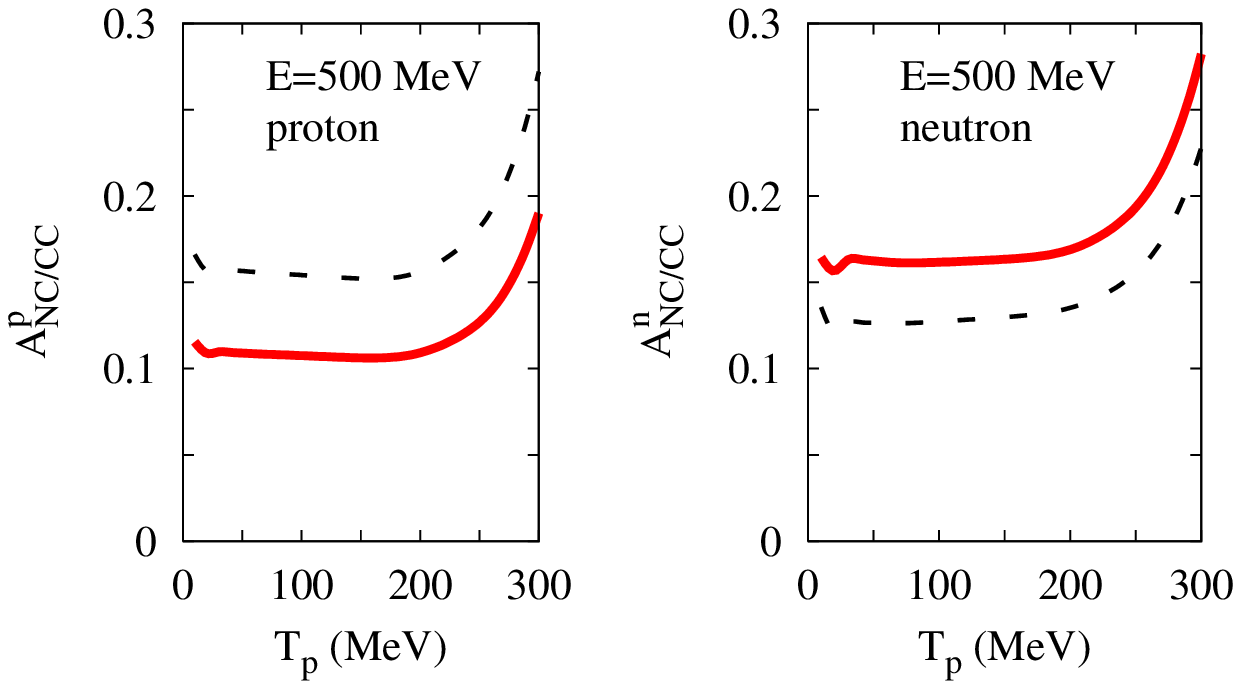}
\caption{(Color online) Asymmetries $A_{NC/CC}^{p}$ and
$A_{NC/CC}^{n}$, Eq.(16), between the NC and CC reactions for an
incident $E_{\nu ({\bar \nu})}$ = 500 MeV. They are calculated for
$g_A^s = -0.19$ (dashed curve) and 0.0 cases (solid curve),
respectively. Upper panel is for the nucleon, while lower panel is
for $^{12}C$. } \label{asym}
\end{figure}

Here two interesting remarks are possible. One is that
$A_{NC/CC}^{p(n)}$ could be a constant if there would be no
strangeness effects. The second point is related to the $\mu_s$
sign. If $g_A^s$ and $G_M^s$ have different $(\pm)$ sign, the
values of $A_{NC/CC}^{p}$ and $A_{NC/CC}^{n}$ become constants
0.12 and 0.16, respectively, because the last two terms in Eq.
(16) are nearly cancelled out. Any deviations from these constants
would imply that both signs are same. For instance, we presented
the case of $g_A^s = -0.19$ and $\mu_s = -0.4$ showing such a
tendency at the upper panels in Fig. \ref{asym}. Both nucleon and
nuclei cases show nearly the same trend. It means that $g_A^s$
effect is nearly evadable from any observable suggested until now.
Divergence of high $T_P$ in the nuclear case of Fig. \ref{asym}
also comes from the CC reaction in the denominator.

\begin{figure}
\includegraphics[width=0.4\linewidth]{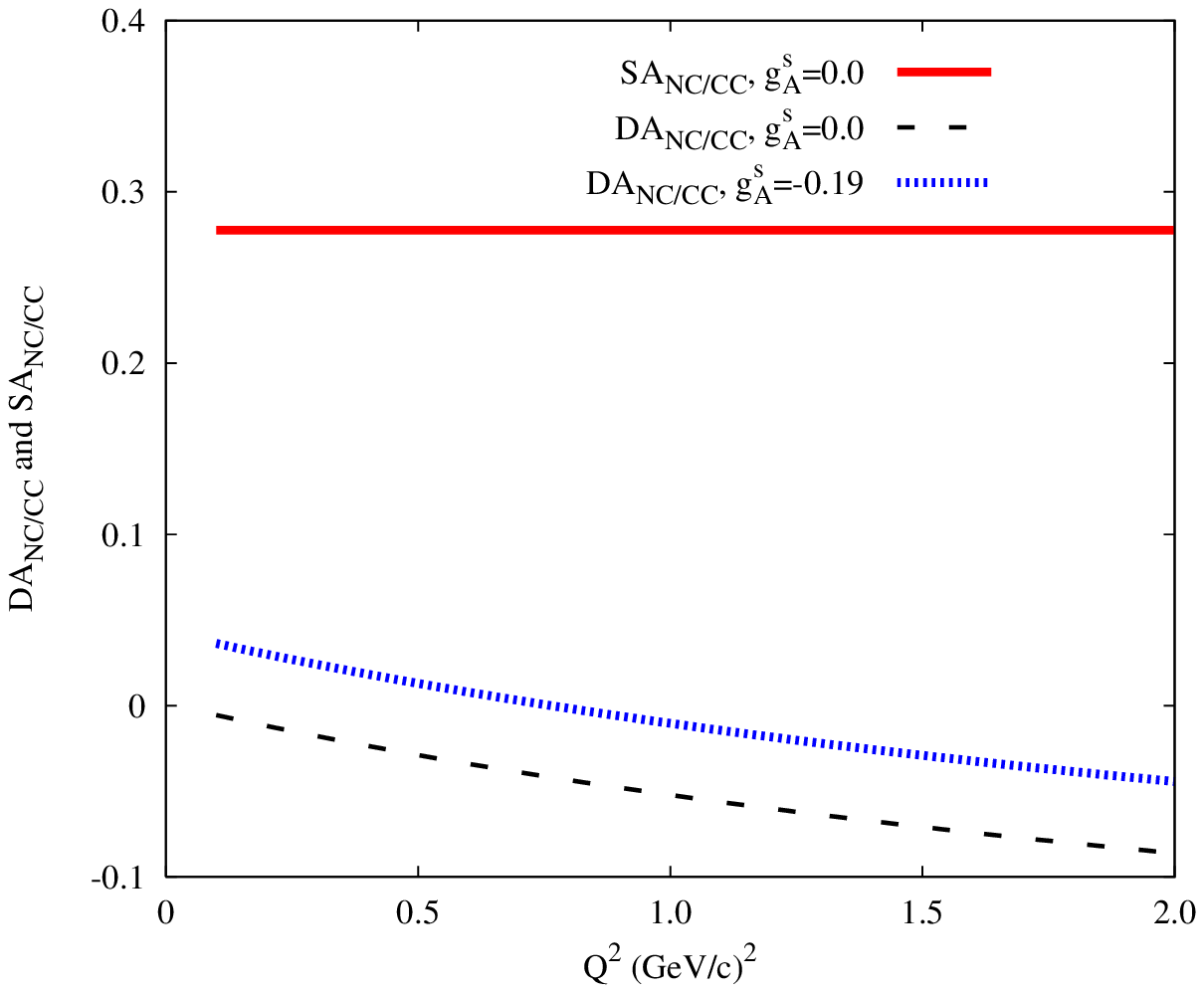}
\includegraphics[width=0.6\linewidth]{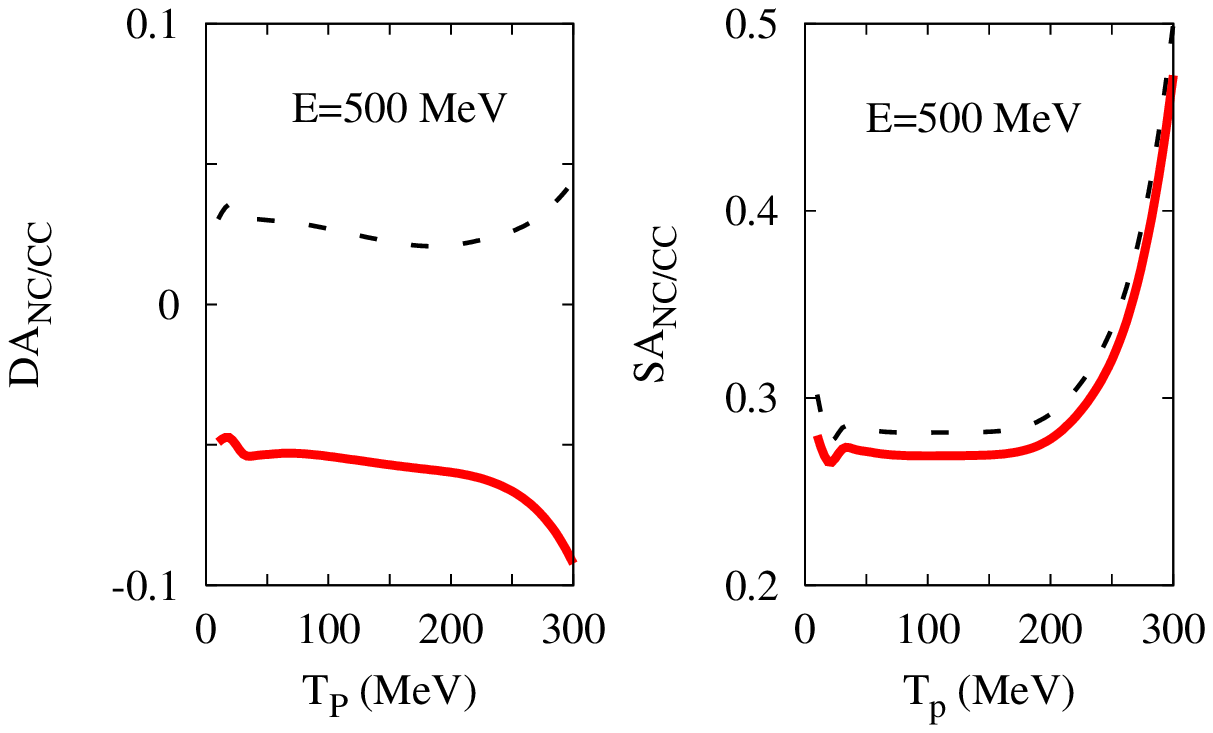}
\caption{(Color online) The differences and summations of the
asymmetries between the NC and CC cross sections of ${\nu - N}$
(upper part) and ${\nu - A}$ scattering for $^{12}$C target (lower
parts) as a function of the knocked-out nucleon kinetic energy.
Solid (red) curves represent the results with $g_A^s=0.0$, dashed
(black) lines are with $g_A^s=-0.19$.} \label{dif-sum}
\end{figure}

As another way to study the strangeness, we introduce the
difference and the sum of asymmetries
\begin{eqnarray}
DA_{NC/CC} & = & A_{NC/CC}^p - A_{NC/CC}^n \simeq - 0.04 -
0.28 ({ {g_A^s } \over { g_A }} + { {G_M^s } \over {G_M^3  }})~,\\
\nonumber SA_{NC/CC} & = & A_{NC/CC}^p + A_{NC/CC}^n = 0.28 + 0.04
{ {g_A^s } \over { g_A }} ~ .
\end{eqnarray}
The sum of asymmetry in Eq. (17) is given only in terms of the
axial part. But the second term is very small by the factor 0.04,
so that the SA is nearly independent of the strangeness, but the
DA depends strongly on the axial strangeness. Figure \ref{dif-sum}
represents the calculation for the difference and the summation of
the asymmetries. On the summation, the $g_A^s$ effect is
negligible, so that the summation of the asymmetry behaves a
constant about 0.28 at low and middle kinetic energies,
independently of free or bound nucleons \cite{cheoun08}. On the
difference, the $g_A^s$ effects clearly appear to be larger than
any other observable. If the values of $g_A^s$ and $\mu_s$ have
different signs, the DA would be constant, but it depends on $Q^2$
if they have same signs as in Fig. \ref{dif-sum}.

Finally, we consider the exclusive $\nu - A$ scattering {\it i.e.}
$A ( \nu, \nu^{'} N )$ because we expect that the strangeness
effects via each nucleon process can be survived in the exclusive
reaction without mutual cancellations. Our terminology for the
exclusive reaction is based on the concept in the electron
scattering \cite{kimepja}.

In the laboratory frame, the exclusive differential cross section
is given by the contraction between the lepton tensor and the
hadron tensor \cite{udias,kimepja}
\begin{eqnarray}
{\frac {d^5\sigma} {dE_f d\Omega_f d\Omega_p}} &=& K [v_L R_L +
v_T R_T + v_{TT}R_{TT} \cos 2\phi_p + v_{LT} R_{LT} \cos \phi_p
\nonumber \\
&+& h ( v'_T R'_T + v'_{LT} R'_{LT} \sin \phi_p )], \label{cs}
\end{eqnarray}
where $\phi_p$ denotes the azimuthal angle of the knocked-out
proton as measured with respect to the incident $\nu$ scattering
plane ($\bf x - \bf z$ plane). $R_L$, $R_T$, $R_{TT}$, $R_{LT}$,
$R'_T$, and $R'_{LT}$ are called the longitudinal, transverse,
transverse-transverse, longitudinal-transverse, polarized
transverse, and polarized longitudinal-transverse response
functions, respectively. The kinematics factor $K$ is given by
$K={\frac {M_N M_{A-1} p} {(2\pi)^3 M_A }} f^{-1}_{rec}
\sigma^Z_M$.

\begin{figure}
\includegraphics[width=0.85\linewidth]{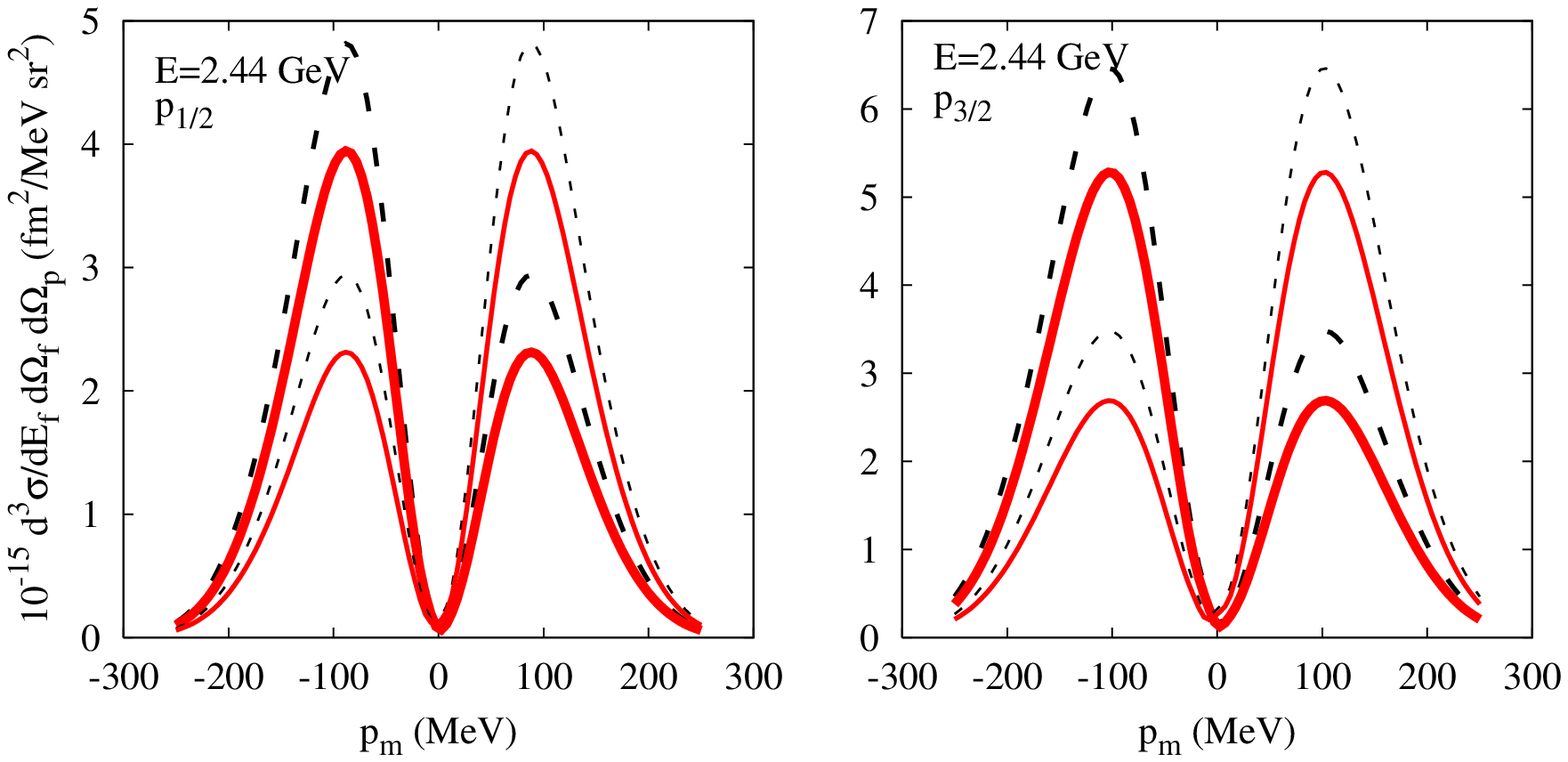}
\caption{(Color online) Neutral current $^{16}$O($\nu, \nu' p$)
cross sections as a function of the missing momentum at
$E_{\nu}=2.4$ GeV. Solid (red) curves are the results for
$g_A^s=0.0$ and dashed (black) lines are for $g_A^s=-0.19$,
respectively. Thick and thin lines are the results for the
incident $\nu$ and ${\bar \nu}$. States of the outgoing nucleon,
$p_{1/2}$ and $ p_{3/2}$, are indicated, respectively.}
\label{exclusive}
\end{figure}

Our results in Fig. \ref{exclusive}, where incident $E_{\nu}= 2.4$
GeV is adopted as a JLAB type's electron energy, show clearly the
strangeness effects, as expected. However, in the experimental
side, this reaction may have actual difficulties in detecting
incident and outgoing neutrinos. However, if one integrates the
scattering angle and averages over the neutrino energies, the five
folded cross section in Eq. (18) becomes a two folded form, that
is, the cross section is going to be the angular distribution of
the knocked-out nucleon by detecting only solid angles of the
knocked-out nucleon.

In conclusion, the $g_A^s$ effect in the elastic $\nu ({\bar
\nu})- N$ scattering for the NC reaction enhances the cross
section for the proton, but reduces it for the neutron. But for
the $\nu - A$ scattering in the QE region, the contributions of
both protons and neutrons compensate each other in the cross
section because of the summation over all nucleons, so that the
net effects on the cross sections are nearly indiscernible due to
the possible cancellations. However the asymmetries between
incident $\nu$ and ${\bar \nu}$ divided by the CC reaction could
be a prominent signal for the strangeness effect. In specific,
differences of the asymmetries could be a feasible quantity for
the signal.

Finally the exclusive reaction like $^{12}$C$( \nu ({\bar \nu}) ,
\nu' ({\bar \nu^{'}}) N )$ could be more efficient tests for the
effect rather than the inclusive reaction $^{12}$C$( \nu ({\bar
\nu}) , \nu' ({\bar \nu^{'}}) )$, because there are no
competitions of the $g_A^s$ effects for the knocked-out protons
and neutrons in the exclusive reaction. Our results for the
exclusive reaction justify clearly this conjecture.

\section*{Acknowledgements}
This work was supported by the Korea Science and Engineering
Foundation(KOSEF) grant funded by the Korea government(MOST)
(R01-2007-000-20569) and one of author, Cheoun, was supported by
the Soongsil University Research Fund.

\newpage

\begin{figure}
\vspace{2.5cm}
\includegraphics[width=1.0\linewidth]{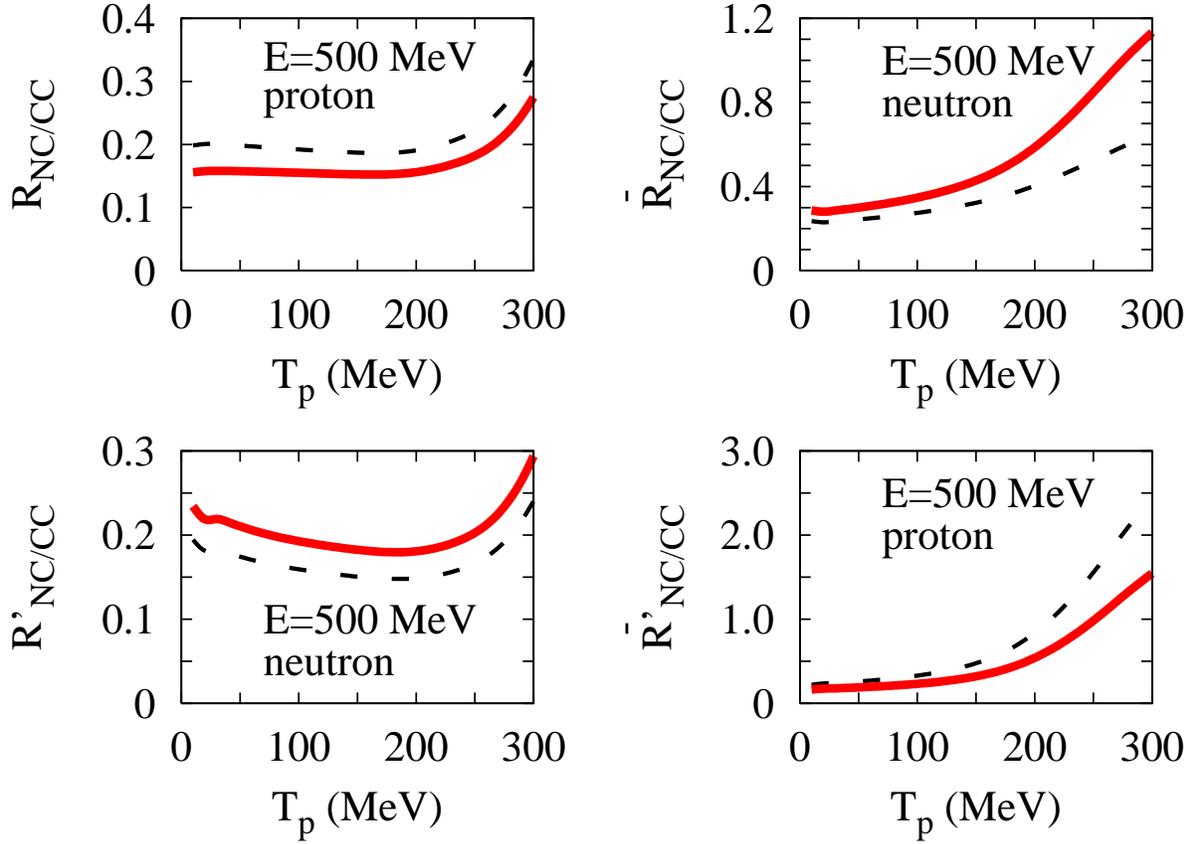}
\caption{(Color online) Ratios of the NC to the CC cross sections
of the $\nu - A$ scattering for $^{12}$C as a function of the
knocked-out nucleon kinetic energy. For the NC reaction, solid
(red) curves represent the results with $g_A^s=0.0$, dashed
(black) lines are with $g_A^s=-0.19$.} \label{ratio}
\end{figure}

\begin{references}

\bibitem{emc89} J. Ashman {\it et al.}, EMC Collaboration, Nucl.
Phys. {\bf B 328}, 1 (1989).
\bibitem{Albe02} W. M. Alberico, S. M. Bilenky, and C. Maieron, Phys.
Rep. {\bf 358}, 227, (2002).
\bibitem{garvey}G. T. Garvey, S. Krewald, E. Kolbe, and K.
Langanke, Phys. Rev. C {\bf 48}, 1919 (1993); Phys. Lett. B {\bf
289}, 249 (1992).

\bibitem{Young06} R. D. Young, J. Roche, R. D. Carlini, and A. W
Thomas, Phys. Rev. Lett., {\bf 97}, 102002, (2006).

\bibitem{Pate05} Stephen F. Pate, Eur. Phys. J. A {\bf 24}, s2,
67, (2005).
\bibitem{Vent06}B. I. S. van der Ventel, and J. Piekarewicz, Phys.
Rev. {\bf C 69}, 035501, (2004).
\bibitem{cheoun08}Myung-Ki Cheoun and K. S. Kim, J. Phys. {\bf G 35}, 065107, (2008).

\bibitem{kimplb}K. S. Kim, Myung Ki Cheoun, and B. G. Yu, to be appeared,
Phys. Rev {\bf C}, arXiv:0707.2767 (2008).

\bibitem{umino}Y. Umino, J. M. Udias, Phys. Rev. C {\bf 52}, 3399
(1995); Y. Umino, J. M. Udias, and P. J. Mulders, Phys. Rev. Lett.
{\bf 74}, 4993 (1995).


\bibitem{giusti1}Andrea Meucci, Carlotta Giusti, and Franco Davide
Pacati, Nucl. Phys. {\bf A739}, 277 (2004); Nucl. Phys. {\bf
A744}, 307 (2004); Nucl. Phys. {\bf A773}, 250(2006).

\bibitem{udias}M. C. Martinez, P. Lava, N. Jahowicz, J.
Ryckebusch, K, Vantournhout, and J. M. Udias, Phys. Rev. C {\bf
73}, 024607 (2006).

\bibitem{clark}E. D. Cooper, S. Hama, B. C. Clark, and R. L.
Mercer, Phys. Rev. C {\bf 47}, 297 (1993).


\bibitem{kim07}K. S. Kim, B. G. Yu, M. K. Cheoun, T. K. Choi, and M. T. Cheon,
J. Phys. G {\bf 34}, 2643 (2007).

\bibitem{madrid}J. M. Udias, P. Sarriguren, E. Moya de Guerra, E.
Garrido, and J. A. Caballero, Phys. Rev. C {\bf 53}, R1488 (1996).





\bibitem{kimepja}K. S. Kim, Myung Ki Cheoun, Yeungun Chung, and
Hyung Joo Nam, Eur. Phys. J. A {\bf 11}, 147 (2001).


\end{references}
\end{document}